# FENN: Feature-enhanced neural network for solving partial differential equations involving fluid mechanics

Jiahao Song[1,2,3], Wenbo Cao[1,2,3], Weiwei Zhang[1,2,3,*]

[1] School of Aeronautics, Northwestern Polytechnical University, Xi'an 710072, China
[2] International Joint Institute of Artificial Intelligence on Fluid Mechanics, Northwestern Polytechnical University, Xi'an, 710072, China
[3] National Key Laboratory of Aircraft Configuration Design, Xi'an 710072, China
* Corresponding author. E-mail: aeroelastic@nwpu.edu.cn

**Abstract** Physics-informed neural networks (PINNs) have shown remarkable prospects in solving forward and inverse problems involving partial differential equations (PDEs). However, PINNs still face the challenge of high computational cost in solving strongly nonlinear PDEs involving fluid dynamics. In this study, inspired by the input design in surrogate modeling, we propose a feature-enhanced neural network. By introducing geometric features including distance and angle or physical features including the solution of the potential flow equation in the inputs of PINNs, FENN can more easily learn the flow, resulting in better performance in terms of both accuracy and efficiency. We establish the feature networks in advance to avoid the invalid PDE loss in FENN caused by neglecting the partial derivatives of the features with respect to space-time coordinates. Through five numerical experiments involving forward, inverse, and parametric problems, we verify that FENN generally reduces the computational cost of PINNs by approximately four times. In addition, the numerical experiments also demonstrate that the proposed method can reduce the number of observed data for inverse problem and successfully solve the parametric problem where PINNs fail.

## 1 Intruduction

Solving partial differential equations (PDEs) involving fluid mechanics is a critical task in many engineering fields, including aerospace, weather forecasting, and natural gas transportation. Obtaining solutions accurately and efficiently is essential to ensuring the reliability and effectiveness of engineering. Over the past few decades, researchers have developed various computational fluid dynamics (CFD) methods to solve PDEs, such as the finite volume method [1], finite difference method [2], and finite element method [3]. These methods are well-established for solving forward problems, where the governing equations and boundary conditions are known. In recent years, with the rapid development of computer technology and machine learning, numerous methods for solving PDEs based on deep neural networks have emerged, with physics-informed neural networks (PINNs) [4] being a typical representative. The central idea of PINNs is integrating the PDE residual into the loss function of the neural networks, ensuring that the networks minimize residual while approaching the boundary conditions or observed data. Compared to CFD methods, PINNs offer several advantages: First, PINNs utilize automatic differentiation (AD) [5] to compute partial derivatives, enabling a meshfree method that eliminates the need for mesh generation. Second, PINNs can easily integrate various observed data to solve inverse problems, such as reconstructing velocity and pressure fields from concentration field [6]. Third, PINNs can solve parametric problems [7-9], obtaining the flow fields for all flow conditions in the predefined parameter space with a single training, while CFD methods can only solve one flow condition at a time.

The unique advantages of PINNs have prompted researchers to apply them to solving PDEs involving fluid mechanics. Raissi et al. [6] used this method to solve inverse problems, reconstructing the velocity and pressure fields based on the concentration field. Jin et al. [10] introduced a solver NSFnets for the incompressible Navier-Stokes equations and demonstrated its effectiveness in a wide range of problems. Mao et al. [11] and Jagtap et al. [12] applied PINNs to solve forward and inverse problems involving supersonic flows. Song et al. [13] proposed volume weighting PINNs to eliminate the ill-conditioning of the PDE loss caused by nonuniform collocation points, they assigned to the PDE residual at each collocation point a weight, which is the volume occupied by that point in the computational domain. Cao et al. [14] developed an efficient solver NNfoil by combining mesh transformation with PINNs for inviscid flow around airfoils. Furthermore, they extended this method to solve the parametric problem about the complete state-space, which includes space-time coordinates, Mach number, angle of attack and shape parameters [8]. Wu et al. [15] enhanced PINNs in thin-layer flows by employing variable linear transformation. PINNs have also been used to solve turbulence flow [16-18], complex three-dimensional flow [19, 20], and fluid-structure interaction problems [21, 22]. These studies have made significant contributions to the application of PINNs in fluid mechanics. However, they generally focus on improving the accuracy and generality of PINNs, without paying attention to accelerating convergence. Fast flow simulation is beneficial for improving engineering efficiency and reducing computational cost, becoming an important branch of CFD [23-25]. Therefore, this study aims to propose a method to accelerate the convergence of PINNs for flow simulation.

Our inspiration comes from input design in surrogate modeling studies [26, 27]. Unlike PINNs, which use space-time coordinates as network inputs, researchers working on surrogate modeling generally design specialized input features to improve modeling accuracy and generalization. Some representative studies include the signed distance function (SDF) [28-30], metrics [31-33], and feature extraction [34-36]. Deng et al. [30] demonstrated the advantages of SDF over space coordinates in turbulence modeling for supercritical airfoils. Hu and Zhang [37] developed a subsonic flow model for airfoils using features such as distance and angle as inputs, achieving remarkable performance. Zuo et al. [34] utilized Transformer to extract geometric information from airfoils as model inputs, improving interpretability while ensuring both accuracy and generalization. These studies have demonstrated that well-designed features can enhance the performance of neural networks. In this study, we propose a feature-enhanced neural network (FENN) to accelerate the convergence of PINNs in fluid dynamics by introducing specifically designed geometric or physical features in the inputs of PINNs. We establish the feature networks in advance to ensure the partial derivatives in PDEs are correctly computed during FENN training. We validate the effectiveness of FENN by solving forward, inverse, and parametric problems under various flow conditions and geometries.

The remainder of the paper is organized as follows. We first introduce the methodology in Section 2, including an overview of PINNs, the proposed FENN and feature design. The strategy to ensure the correct partial derivatives is also presented. In Section 3, we validate the effectiveness of FENN through five numerical experiments. Finally, Section 4 provides concluding remarks and suggests directions for future research.

## 2 Methodology

2.1 Physics-informed neural networks

PINNs employ deep neural networks to solve PDEs based on optimization algorithms. The network takes the space-time coordinates $(\mathbf{x}, t)$ as the inputs, and the output representing the solution $u(\mathbf{x}, t)$ of the PDEs. We consider the PDE defined on a domain $\Omega \subset \mathbb{R}^d$:

$$\begin{cases} u_t + \mathcal{N}[u] = 0, & \mathbf{x} = (x_1,...,x_d) \in \Omega, t \in [0,T] \\ u(\mathbf{x},0) = g(\mathbf{x}), & \mathbf{x} \in \Omega \\ \mathcal{B}[u] = 0, & \mathbf{x} \in \partial\Omega, t \in [0,T] \end{cases} \quad (1)$$

where $u$ denotes the solution of the PDE, $\mathcal{N}[\cdot]$ represents the general differential operator that includes both linear and nonlinear terms, and $\mathcal{B}[\cdot]$ is the boundary condition operator. The initial condition is defined by $g(\mathbf{x})$.

The loss function of PINNs is defined as

$$\mathcal{L} = \mathcal{L}_{ic} + \mathcal{L}_{bc} + \mathcal{L}_r + \mathcal{L}_{da} \quad (2)$$

where

$$\mathcal{L}_{ic} = \frac{1}{N_{ic}} \sum_{i=1}^{N_{ic}} \left| u_\theta(\mathbf{x}_{ic}^i, 0) - g(\mathbf{x}_{ic}^i) \right|^2 \quad (3)$$

$$\mathcal{L}_{bc} = \frac{1}{N_{bc}} \sum_{i=1}^{N_{bc}} \left| \mathcal{B}[u_\theta(\mathbf{x}_{bc}^i, t_{bc}^i)] \right|^2 \quad (4)$$

$$\mathcal{L}_r = \frac{1}{N_r} \sum_{i=1}^{N_r} \left| \frac{\partial u_\theta}{\partial t}(\mathbf{x}_r^i, t_r^i) + \mathcal{N}[u_\theta(\mathbf{x}_r^i, t_r^i)] \right|^2 \quad (5)$$

$$\mathcal{L}_{da} = \frac{1}{N_{da}} \sum_{i=1}^{N_{da}} \left| u_\theta(\mathbf{x}_{da}^i, t_{da}^i) - u(\mathbf{x}_{da}^i, t_{da}^i) \right|^2 \quad (6)$$

In Eqs. (3)-(6), $\theta$ represent the network parameters, $u_\theta(\mathbf{x},t)$ denotes the output of PINNs. $\{\mathbf{x}_{ic}^i, 0\}_{i=1}^{N_{ic}}$ and $\{\mathbf{x}_{bc}^i, t_{bc}^i\}_{i=1}^{N_{bc}}$ are training points for the initial condition and boundary condition, the corresponding initial condition loss $\mathcal{L}_{ic}$ and boundary loss $\mathcal{L}_{bc}$ are typically used only for solving forward problems. $\{\mathbf{x}_r^i, t_r^i\}_{i=1}^{N_r}$ are collocation points used to evaluate the PDE loss $\mathcal{L}_r$. $\{\mathbf{x}_{da}^i, t_{da}^i\}_{i=1}^{N_{da}}$ is observed data points, $u(\mathbf{x}_{da}^i, t_{da}^i)$ represents available observed data, and the corresponding data loss $\mathcal{L}_{da}$ only exists in solving inverse problems. The partial derivatives in the loss function are computed using automatic differentiation. During the network training, $\mathcal{L}_{ic}$, $\mathcal{L}_{bc}$ and $\mathcal{L}_{da}$ ensure the network output approach the initial condition, boundary condition and observed data, while $\mathcal{L}_r$ constrains it to satisfy the PDE. The conventional PINNs framework is illustrated in Fig. 1. It is noted that in this study, we employ volume weighting PDE loss (7) proposed by Song et al. [13] to replace Eq. (5), the former has been proven to perform better on nonuniform collocation points. In Eq. (7), $V(\mathbf{x}_r^i, t_r^i)$ represents the volume occupied by $(\mathbf{x}_r^i, t_r^i)$ in the computational domain.

$$\mathcal{L}_r = \sum_{i=1}^{N_r} \left| \{ \frac{\partial u_\theta}{\partial t}(\mathbf{x}_r^i, t_r^i) + \mathcal{N}[u_\theta(\mathbf{x}_r^i, t_r^i)] \} V(\mathbf{x}_r^i, t_r^i) \right|^2 / \sum_{i=1}^{N_r} [V(\mathbf{x}_r^i, t_r^i)]^2 \qquad (7)$$

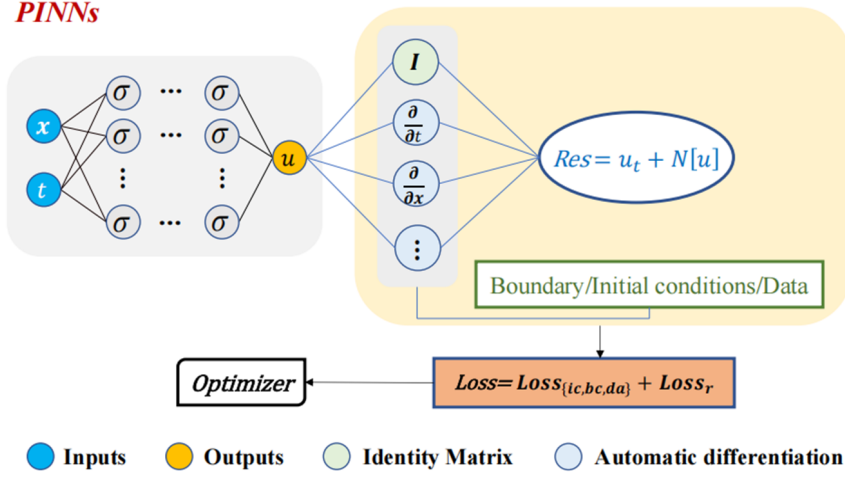

Figure 1. A schematic of PINNs for solving partial differential equations.

2.2 Feature-enhanced neural network

In surrogate modeling, the input features of the network significantly impact model performance [28, 34, 38]. Input features such as the signed distance function [29, 30] and metrics [31, 32] have been proven to enhance model accuracy and generalization. In contrast, the inputs of PINNs only include space-time coordinates. Inspired by these studies, we propose a novel method for solving PDEs involving fluid mechanics, called the feature-enhanced neural network (FENN). Specifically, we combine space-time coordinates $(\mathbf{x}, t)$ with designed features $F(\mathbf{x}, t)$ as inputs to the neural network, while keeping other settings the same as in PINNs. By introducing features in the network inputs, FENN can learn the flow more easily than PINNs, resulting in better performance in both accuracy and efficiency. Since FENN only adds a few neurons to the network inputs, it leads to almost no additional computational cost. The feature design will be discussed in detail in Section 2.3.

It is important to note that simply introducing features in the inputs of the FENN is insufficient. Since $F$ are functions of $(\mathbf{x}, t)$. $\partial F / \partial \mathbf{x}$ and $\partial F / \partial t$ should be considered when using automatic differentiation to compute the partial derivatives in the PDEs. Taking the computation of $\partial u_\theta / \partial x$ as an example. According to the chain rule, the correct $\partial u_\theta / \partial x$ can be obtained by

$$\frac{\partial u_\theta}{\partial x} = (\frac{\partial u_\theta}{\partial x})_s + \sum_{j=1}^{N} \frac{\partial u_\theta}{\partial F_j} \frac{\partial F_j}{\partial x}, \quad j = 1, ..., N \qquad (8)$$

where $N$ represents the number of features. $(\partial u_\theta / \partial x)_s$ and $\partial u_\theta / \partial F_j$ represent partial derivatives of the output $u_\theta$ of FENN with respect to its inputs $x$ and $F_j$, which are computed by automatic differentiation. Simply introducing $F_j$ to the network inputs results in $\partial F_j / \partial x = 0$, which in turn leads to an incorrect $\partial u_\theta / \partial x$. To address this issue, we train $N$ feature networks offline in advance to obtain the regression models of $F_j$ with respect to $(\mathbf{x},t)$. These models are then employed as features in the FENN inputs, as shown in Figure 2. The loss function of the feature networks is

$$\mathcal{L}_{F_j} = \frac{1}{M}\sum_{i=1}^{M}\left|F_j^\theta(\mathbf{x}_i,t_i) - F_j^{la}(\mathbf{x}_i,t_i)\right|^2, \quad j=1,...,N \tag{9}$$

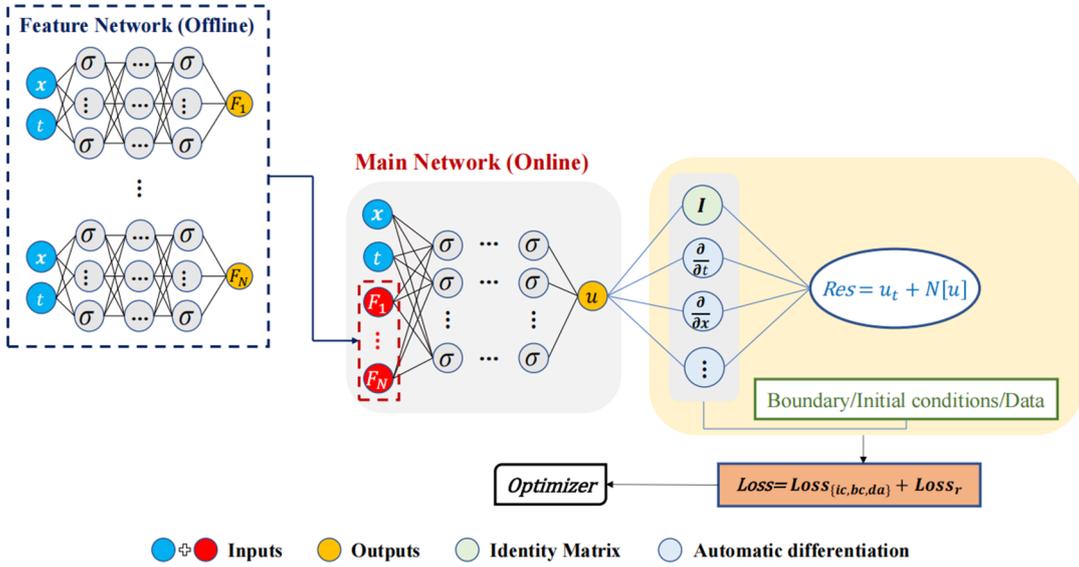

Figure 2. A schematic of FENN for solving partial differential equations.

where $F_j^\theta(\mathbf{x},t)$ represent the output of the feature networks, $F_j^{la}$ represent the label data. $\partial F_j^\theta / \partial \mathbf{x}$ and $\partial F_j^\theta / \partial t$ are computed in the feature networks by automatic differentiation during the main network training. We further avoid the additional computation originating from the feature networks at each epoch of the main network by computing $\partial F_j^\theta / \partial \mathbf{x}$ and $\partial F_j^\theta / \partial t$ before training the main network and reconstructing features based on them, as elaborated in the Appendix A. It should be noted that if an analytical expression of $F_j$ with respect to $(\mathbf{x},t)$ exists, we use it to replace the corresponding feature network.

Compared to PINNs, training the feature networks leads to additional computational cost in FENN. However, compared to the main network, the feature networks have fewer network parameters and are easier to train because they are supervised. Therefore, their computational cost is much lower than that of the main network. Detailed results from numerical experiments are provided in Section 3.

2.3 Feature design

We first refer to the input features designed by Hu et al. [37] in surrogate modeling, designing two features: distance and angle, as shown in Figure 3. Specifically, for point $\mathbf{x}$, the distance $D$ is the shortest distance from $\mathbf{x}$ to the wall (the corresponding point $\mathbf{x}_l$ on the wall is defined as the local point). The angle $\Phi$ is the angle $\varphi$ between the tangent direction of $\mathbf{x}_l$ and the wall $0°$ angle of attack, multiplied by the distance attenuation function $\delta(D)$

$$\delta(D) = 1 - \frac{D^3}{D^3 + c/10} \tag{10}$$

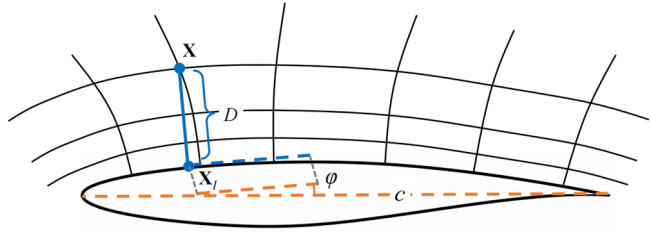

Figure 3. A schematic of the distance and angle.

where $c$ represents the characteristic length, such as the diameter of a cylinder or the chord length of an airfoil. The surprisingly high-accuracy results in [37] demonstrate the beneficial effects of these two features. Since both $D$ and $\Phi$ are extracted based on the shape of the wall, we define them as geometric features, which are the main features used in this study.

In addition, We design physical features. In fluid mechanics, the Navier-Stokes equations degenerate into the Euler equations under inviscid conditions, and the Euler equations further degenerate into the potential flow equation under irrotational conditions. Figure 4 shows an example of two-dimensional flow around a circular cylinder. Potential flow contains basic flow properties and has a simpler relationship with inviscid and viscous flows compared to space-time coordinates. Meanwhile, the difficulty and computational cost of solving the potential flow equation are much lower than those of the Euler and Navier-Stokes equations, and it has analytical solution for some problems. Therefore, we design the physical features using the solution of the potential flow equation. Specifically, we introduce the $x$-component $u^{pf}$ of velocity vector, the $y$-component $v^{pf}$ of velocity vector, and the pressure $p^{pf}$ in the inputs of the FENN.

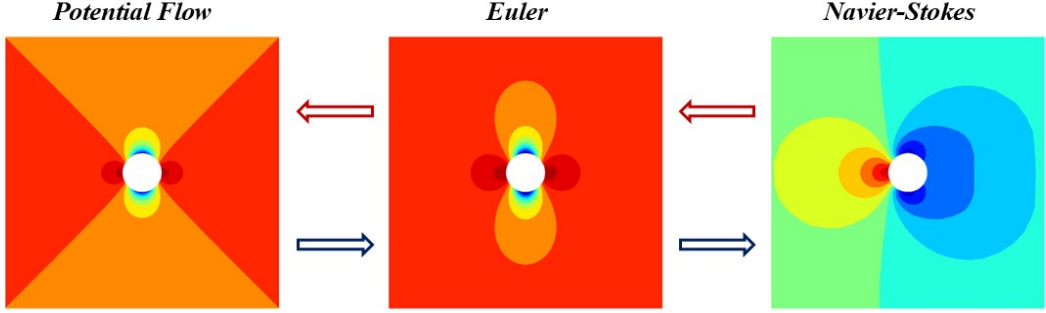

Figure 4. A schematic of Navier Stokes equations degeneration.

## 3  Results

We validate the effectiveness of FENN by solving three forward problems, one inverse problem, and one parametric problem about the angle of attack. The forward problems include: inviscid compressible flow around a circular cylinder, viscous incompressible flow around the NACA0012 airfoil and viscous incompressible flow around a circular cylinder. The inverse problem involves reconstructing high-resolution velocity and pressure fields of viscous incompressible flow around a circular cylinder using sparse velocity data. The parametric problem about the angle of attack involves the viscous incompressible flow around the NACA2412 airfoil. We use the tanh activation function and initialize the network parameters using normal distributions [39]. Limited-memory Broyden-Fletcher-Goldfarb-Shanno (L-BFGS) [40] optimizer is employed to perform gradient descent, with the maximum number of inner iterations per epoch set to 20. The feature networks contain 3 hidden layers with 20 neurons per layer, and they are trained for 1000 epochs. Appendix B summarizes the loss convergence of the feature networks.

To evaluate the accuracy of PINNs and FENN, we compute the relative $L_2$ error

$$\frac{\|U_\theta - U\|_2}{\|U\|_2} \tag{11}$$

where $U_\theta$ represents the predicted solution by the neural network, $U$ represents the reference solution obtained by the finite volume method. The study develops programs based on the PyTorch platform and executes the algorithm on an NVIDIA GeForce RTX 4090 GPU.

3.1  Inviscid compressible flow around a circular cylinder

We first solve the inviscid compressible flow around a circular cylinder, which is a common benchmark in fluid mechanics for method validation. The dimensionless governing equations for this problem are

$$\rho(\frac{\partial u}{\partial x}+\frac{\partial v}{\partial y})+u\frac{\partial \rho}{\partial x}+v\frac{\partial \rho}{\partial y}=0$$

$$\rho(u\frac{\partial u}{\partial x}+v\frac{\partial u}{\partial y})+\frac{\partial p}{\partial x}=0 \quad (12)$$

$$\rho(u\frac{\partial v}{\partial x}+v\frac{\partial v}{\partial y})+\frac{\partial p}{\partial y}=0$$

$$\frac{\rho}{\gamma(\gamma-1)}(u\frac{\partial T}{\partial x}+v\frac{\partial T}{\partial y})+Ma^2 p(\frac{\partial u}{\partial x}+\frac{\partial v}{\partial y})=0$$

where $u$ denotes the $x$-component of the velocity vector $\mathbf{V}$, $v$ the $y$-component. $\rho$, $T$ and $p$ represent density, temperature and pressure, respectively, satisfying the relationship $p=\rho T/(\gamma Ma^2)$. $Ma$ represent the Mach number. $\gamma=1.4$ is the specific heat ratio. We set $Ma=0.4$.

Figure 5 shows the computational domain and the distribution of collocation points. The cylinder is placed at $(x,y)=(0,0)$ with a diameter of $d=1$, while the far field is positioned at the same center with a diameter of 40. The number of collocation points is $N_r=4800$. The boundary conditions $[\rho_\infty,u_\infty,v_\infty,T_\infty]=[1,1,0,1]$ and $\mathbf{V}\cdot\mathbf{n}=0$ are enforced to the far field and the cylinder ($\mathbf{n}$ represent a unit vector normal to the cylinder), respectively, with the number of training points $N_{bc}=80$ for both. We solve the problem using PINNs and FENN, respectively. The inputs of PINNs are $[x,y]$, the outputs are $[\rho,u,v,T]$. The inputs of FENN are $[x,y]$ combined with the distance $D$ and angle $\Phi$, forming the set $[x,y,D,\Phi]$, the outputs are the same as PINNs. Both networks contain 5 hidden layers with 64 neurons per layer. PINNs and FENN are trained for 2000 and 400 epochs, respectively. It should be noted that since the distance under the cylinder can be expressed analytically as $D=\sqrt{x^2+y^2}-d$, we only establish the feature network for the angle. The test error of the feature network is 8.8e-6, achieving a high-accuracy representation of the angle. Its training time is 0.14 minutes.

Figure 6(a) and (b) show the comparison of the loss convergence between PINNs and FENN. We use the pressure coefficient $C_p=(p-p_\infty)/[0.5\rho_\infty(u_\infty^2+v_\infty^2)]$ on the cylinder to evaluate the accuracy of the solution, the convergence history of the relative $L_2$ errors between the $C_p$ obtained by PINNs, FENN and the reference solution are given in Figure 6(c). We observe that the relative $L_2$ error of FENN trained for 400 iterations and PINNs trained for 2000 iterations is nearly identical, approximately 1%. The corresponding pressure coefficients are shown in Figure 7. The training times for PINNs and FENN are 7.78 minutes and 1.61 minutes, respectively. Adding the training time of the feature network, the total training time for FENN is 1.75 minutes. It is reduced by 4.45 times compared to PINNs.

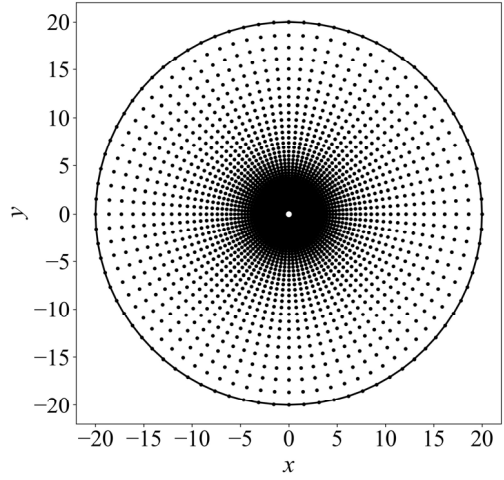

Figure 5. The computational domain and the distribution of collocation points for solving inviscid compressible flow around a circular cylinder.

3.2 Viscous incompressible flow around the NACA0012 airfoil

Here, we consider the viscous incompressible flow around the NACA0012 airfoil. The dimensionless governing equations are

$$\frac{\partial u}{\partial x} + \frac{\partial v}{\partial y} = 0$$

$$u\frac{\partial u}{\partial x} + v\frac{\partial u}{\partial y} + \frac{\partial p}{\partial x} - \frac{1}{Re}(\frac{\partial^2 u}{\partial x^2} + \frac{\partial^2 u}{\partial y^2}) = 0 \quad (13)$$

$$u\frac{\partial v}{\partial x} + v\frac{\partial v}{\partial y} + \frac{\partial p}{\partial y} - \frac{1}{Re}(\frac{\partial^2 v}{\partial x^2} + \frac{\partial^2 v}{\partial y^2}) = 0$$

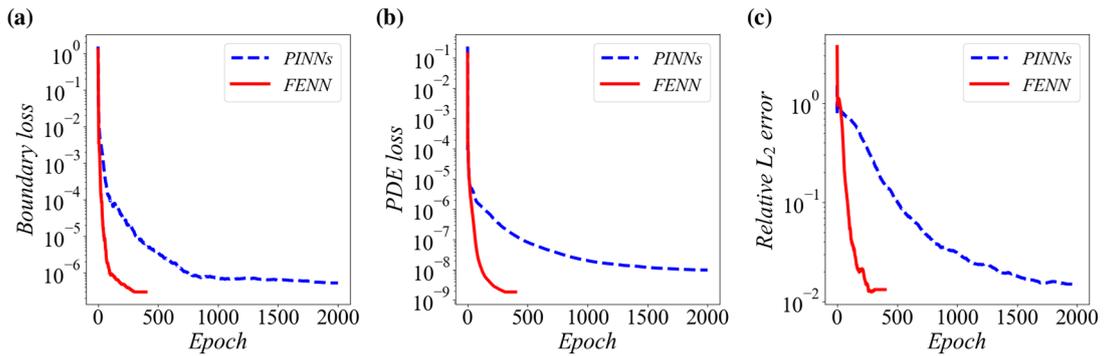

Figure 6. Comparison of the convergence history obtained by PINNs and FENN for solving inviscid compressible flow around a circular cylinder. (a) Boundary loss. (b) PDE loss. (c) Relative $L_2$ error of the pressure coefficient on the cylinder.

where $u$ denotes the $x$-component of the velocity vector, $v$ the $y$-component. $p$ represents pressure, $Re$ represents the Reynolds number. We set $Re = 400$, the angle of attack is $\alpha = 5°$.

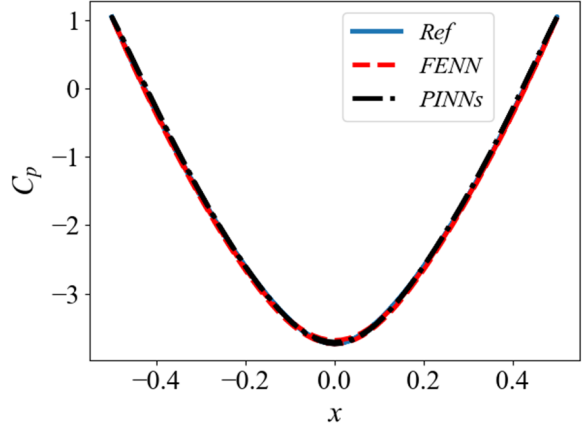

Figure 7. Comparison of the pressure coefficient on the cylinder obtained by finite volume method (blue), PINNs (black), and FENN (red) for solving inviscid compressible flow around a circular cylinder.

The computational domain and the distribution of collocation points are shown in Figure 8. The leading edge point of the airfoil is located at $(x, y) = (0,0)$ with chord length $l = 1$. The number of collocation points is $N_r = 14640$. At the inflow and outflow boundaries, we place $N_{bc} = 200$ and 75 points, and enforce velocity inlet condition $[u_\infty, v_\infty] = [\cos(\alpha), \sin(\alpha)]$ and pressure outlet condition $p_\infty = 0$. At the airfoil, we place $N_{bc} = 200$ points and enforce the no-slip condition $[u, v] = [0, 0]$. We solve the problem using PINNs and FENN, respectively. The inputs of PINNs are $[x, y]$, the outputs are $[u, v, p]$. The inputs of FENN are $[x, y, D, \Phi]$, the outputs are the same as PINNs. Both networks contain 7 hidden layers with 64 neurons per layer. PINNs and FENN are trained for 10000 and 2500 epochs, respectively. In FENN, the testing errors of the distance and angle feature networks are 1.42e-5 and 1.32e-5, with a total training time of 1.31 minutes.

Figure 9 shows the convergence history of PINNs and FENN. We observe that compared to PINNs, FENN obtains more accurate results with only one-quarter of the epochs. Compared to the reference solution, the relative $L_2$ error of the pressure coefficient on the airfoil obtained by PINNs and FENN are 3.6% and 2.8%. The corresponding pressure coefficients are shown in Figure 10. The training times for PINNs and FENN are 51.26 minutes and 13.33 minutes, respectively. Adding the training times of the feature networks, the total training time for FENN is 14.64 minutes, representing a speedup of 3.50 times compared to PINNs.

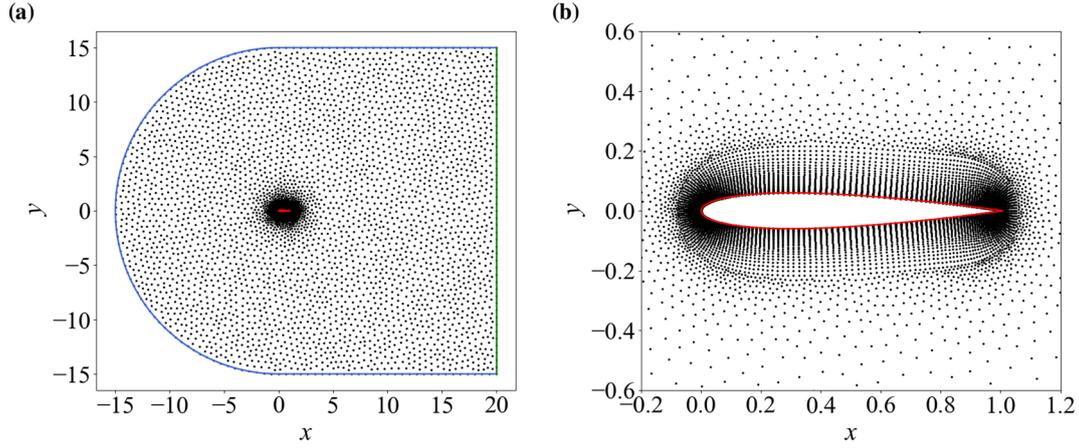

Figure 8. The computational domain and the distribution of collocation points for solving viscous incompressible flow around the NACA0012 airfoil. (a) Entire computational domain. The blue line represents the inflow boundary, the green line represents the outflow boundary, and the red line represents the NACA0012 airfoil. (b) Near the airfoil region.

3.3 Inverse problem involving viscous incompressible flow around a circular cylinder

In this section, we test the performance of FENN in solving inverse problems. Solving inverse problems is a limitation of CFD methods, as they require extremely high computational cost [6]. In contrast, PINNs allow for convenient integration of observed data, demonstrating significant advantages in solving such problems.

We consider reconstructing high-resolution velocity and pressure fields for viscous incompressible flow around a circular cylinder based on sparse velocity data when boundary conditions are unknown, which is a standard inverse problem in fluid mechanics [6, 19]. The dimensionless governing equations are given by Eq. (13) with $Re = 40$.

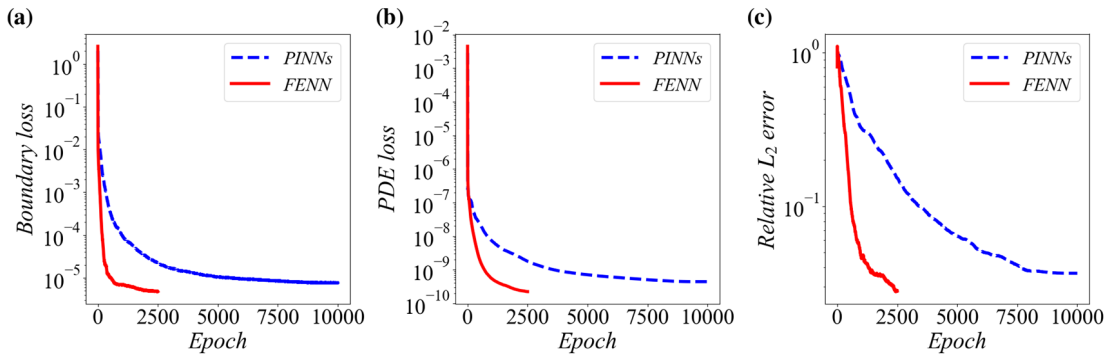

Figure 9. Comparison of the convergence history obtained by PINNs and FENN for solving viscous incompressible flow around the NACA0012 airfoil. (a) Boundary loss. (b) PDE loss. (c) Relative $L_2$ error of the pressure coefficient on the airfoil.

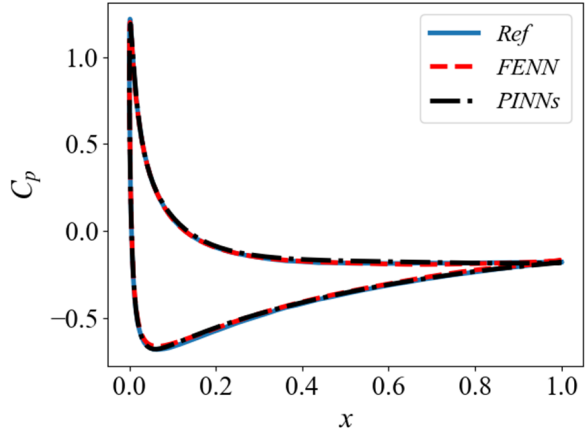

Figure 10. Comparison of the pressure coefficient on the NACA0012 airfoil obtained by finite volume method (blue), PINNs (black), and FENN (red) for solving viscous incompressible flow over the NACA0012 airfoil.

Figure 11(a) shows the computational domain and the distribution of collocation points. The cylinder is placed at $(x, y) = (0,0)$ with a diameter of $d = 1$. The number of collocation points is $N_r = 12300$. According to the fact that the flow field in the near wall region is more complex, we use the probability density of the collocation points to guide the sampling of velocity data, ensuring that the near wall region has denser observed data, as shown in Figure 11(b) (For illustration purpose only, we consider three cases where the number of observed data points is $N_{da} = 20$, 50, and 100). We solve the problem using PINNs and FENN, respectively. The inputs of PINNs are $[x, y]$, the outputs are $[u, v, p]$. The inputs of FENN are $[x, y, D, \Phi]$, the outputs are $[u, v, p]$. Both networks contain 5 hidden layers with 64 neurons per layer. PINNs and FENN are trained for 3000 and 800 epochs, respectively. In FENN, the distance feature is represented by the analytical formula $D = \sqrt{x^2 + y^2} - d$, and the testing error of the angle feature network is 9.69e-6 with the training time of 0.14 minutes.

Table 1 shows the relative $L_2$ errors of the pressure coefficient on the cylinder obtained by PINNs and FENN under three different $N_{da}$. When $N_{da} = 20$, the error of the results obtained by PINNs is 17.59%, while that of FENN is 1.67%. This demonstrates that FENN can significantly improve the reconstruction accuracy of the flow field under extremely sparse data. The point-wise reconstruction errors of the flow field are given in Figure 12. As $N_{da}$ increases to 100, the errors of the results obtained by PINNs and FENN decrease to around 1%. This is expected, as more flow field constraints are incorporated. Figure 13 shows the convergence history of PINNs and FENN under $N_{da} = 20, 100$. We observe that compared to PINNs, FENN generally obtains more accurate results with approximately one-quarter of the epochs. The corresponding pressure coefficients are shown in Figure 14. When $N_{da} = 100$, the training times for PINNs and FENN are 17.01 minutes and 4.67 minutes, respectively. Adding the training time of the angle feature network, the total training time for FENN is 4.81 minutes. Compared to PINNs, the speedup of FENN is 3.54.

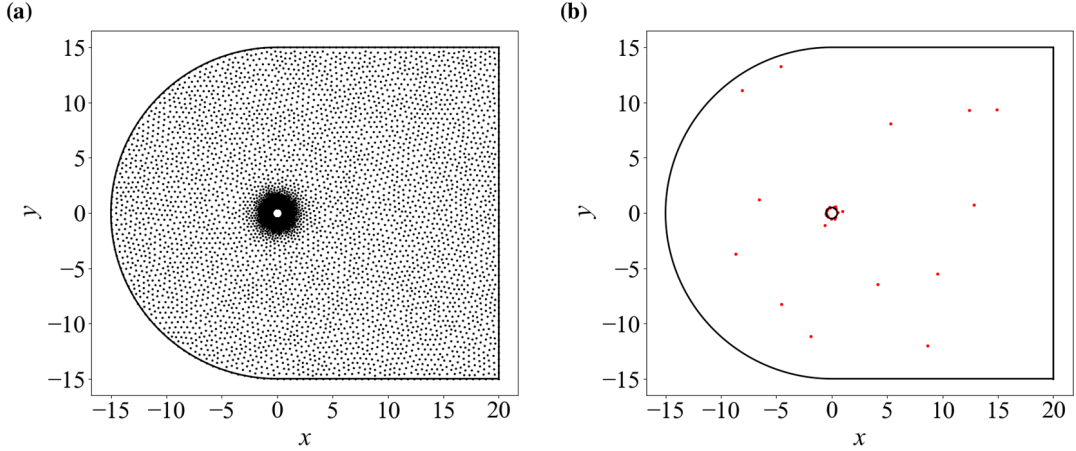

Figure 11. The computational domain and the distribution of sampling points for solving the inverse problem involving viscous incompressible flow around a circular cylinder.

Tabel 1. Relative $L_2$ error of the pressure coefficient on the cylinder at different $N_{da}$.

|  | $N_{da} = 20$ | $N_{da} = 50$ | $N_{da} = 100$ |
| --- | --- | --- | --- |
| PINNs | 17.59% | 2.26% | 1.35% |
| FENN | 1.67% | 1.31% | 1.03% |

3.4 Parametric problem involving viscous incompressible flow around the NACA2412 airfoil

Our next experiment involves solving a parametric problem. PINNs also demonstrate significant advantages over CFD methods for such problems [14, 41]. Taking the parametric problem about angle of attack in fluid mechanics as an example, PINNs can obtain all flow fields within the predefined range of angles of attack by introducing the angle of attack in the network inputs and training once. In contrast, CFD methods can only solve the flow field for a single angle of attack at a time. We consider the parametric problem about the angle of attack involving viscous incompressible flow around the NACA2412 airfoil. The dimensionless governing equations are given by Eq. (13) with $Re = 400$.

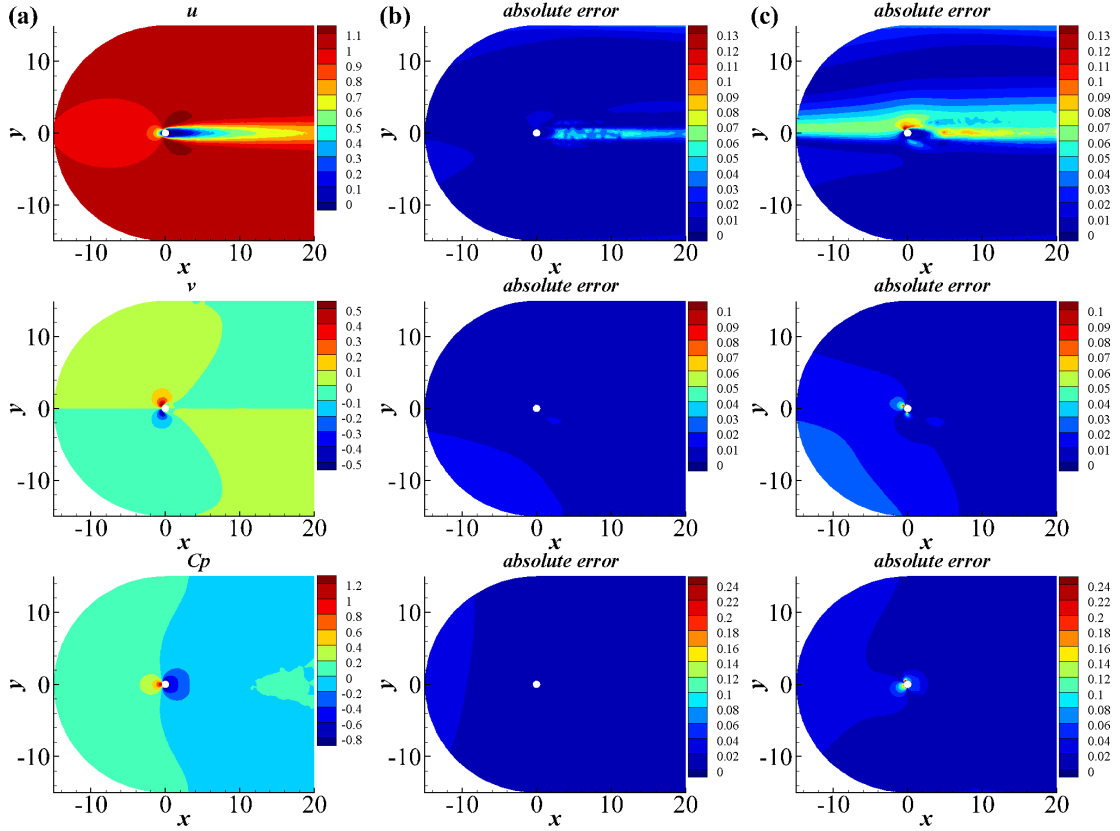

Figure 12. Absolute error of the flow field between (a) the reference solution and the solutions given by (b) FENN and (c) PINNs for solving the inverse problem involving viscous incompressible flow around a circular cylinder at $N_{da} = 20$.

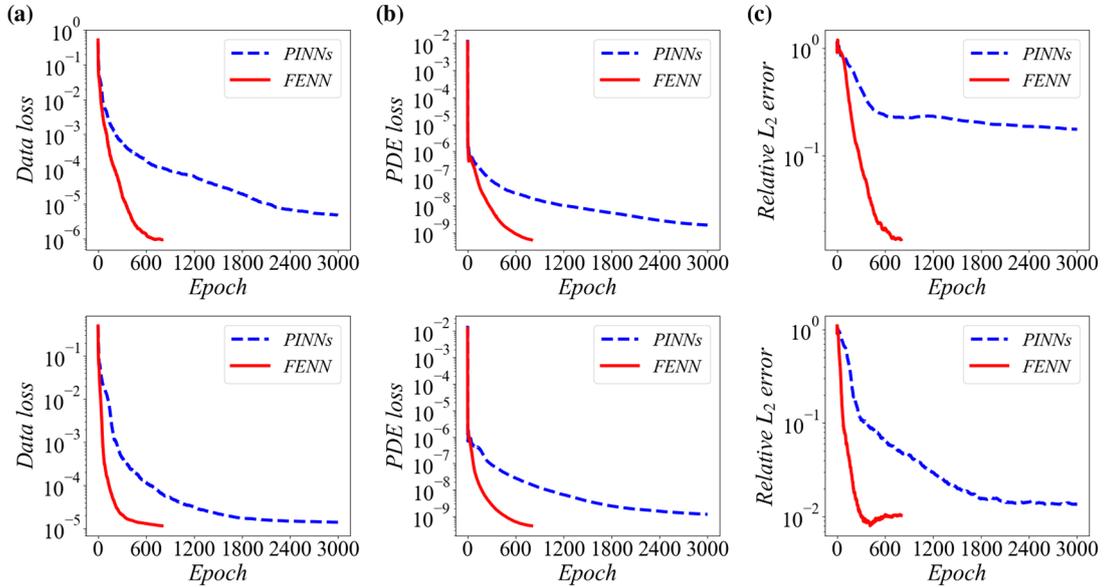

Figure 13. Comparison of the convergence history obtained by PINNs and FENN for solving the inverse problem involving viscous incompressible flow around a circular cylinder at $N_{da} = 20$ (top) and $N_{da} = 100$ (bottom). (a) Data loss. (b) PDE loss. (c) Relative $L_2$ error of the pressure coefficient on the cylinder.

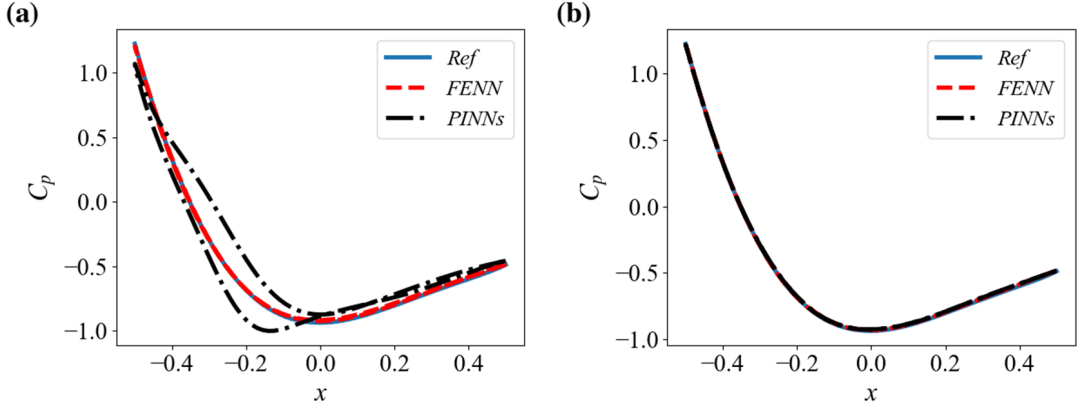

Figure 14. Comparison of the pressure coefficient on the cylinder obtained by finite volume method (blue), PINNs (black), and FENN (red) for solving inverse problem involving viscous incompressible flow around a circular cylinder. (a) $N_{da}=20$. (b) $N_{da}=100$.

The computational domain, the coordinates of the leading edge, and the chord length for this problem are the same as in Section 3.2. The inputs of PINNs are $[x,y,\alpha]$, the outputs are $[u,v,p]$. The inputs of FENN are $[x,y,\alpha,D,\Phi]$, the outputs are the same as PINNs. We set $\alpha \in [-8°, 8°]$. Both networks contain 9 hidden layers with 128 neurons per layer. We utilize the L-BFGS optimizer to perform gradient descent, with the maximum number of inner iterations per epoch set to 1000. PINNs and FENN are trained for 1500 epochs. In each epoch, we resample the collocation points and reinitialize the optimizer. This addresses the incompatibility issue between the LBFGS optimizer and batching, avoiding extensive sampling of collocation points [8]. The number of collocation points is $N_r = 20000$. At the inflow and outflow boundaries, we place $N_{bc}=5000$ and 2000 points, and enforce velocity inlet condition $[u_\infty, v_\infty] = [\cos(\alpha), \sin(\alpha)]$ and pressure outlet condition $p_\infty = 0$. At the airfoil, we place $N_{bc}=4000$ points and enforce the no-slip condition $[u,v]=[0,0]$. In FENN, the testing errors of the distance and angle feature networks are 2.14e-5 and 1.72e-5.

Figure 15 shows the comparison of the loss convergence between PINNs and FENN. We observe the significant convergence acceleration of FENN compared to PINNs. In addition, compared to the solution of a single state problem (Figure 9), the loss of PINNs in the parametric problem is one order of magnitude higher when the slope of the loss curve approaches zero. In contrast, the loss of FENN is of the same order of magnitude as that of the single state problem. We use the pressure coefficient on the airfoil at six different angles of attack to evaluate the accuracy, as shown in Figure 16. We observe that the results obtained by PINNs are unacceptable, while those obtained by FENN are in better agreement with the reference solutions.

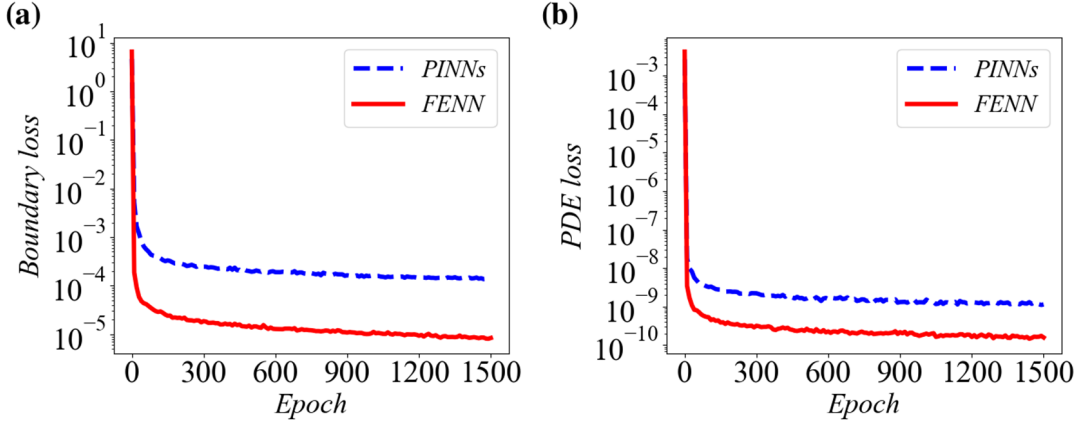

Figure 15. Comparison of the convergence history obtained by PINNs and FENN for solving the parametric problem involving viscous incompressible flow around the NACA2412 airfoil. (a) Boundary loss. (b) PDE loss.

3.5 Forward problem involving viscous incompressible flow around a circular cylinder

Finally, we validate the effectiveness of physical features by solving the forward problem involving viscous incompressible flow around a circular cylinder. The dimensionless governing equations are given by Eq. (13) with $Re = 40$.

The computational domain and the distribution of collocation points are shown in Figure 17. The cylinder is placed at $(x, y) = (0, 0)$ with a diameter of $d = 1$. The number of collocation points is $N_r = 12300$. At the inflow and outflow boundaries, we place $N_{bc} = 200$ and 75 points, and enforce velocity inlet condition $[u_\infty, v_\infty] = [1, 0]$ and pressure outlet condition $p_\infty = 0$. At the cylinder, we place $N_{bc} = 200$ points and enforce the no-slip condition $[u, v] = [0, 0]$. We solve the problem using PINNs and FENN, respectively. The inputs of PINNs are $[x, y]$, the outputs are $[u, v, p]$. The inputs of FENN are $[x, y]$ combined with the solution $[u^{pf}, v^{pf}, p^{pf}]$ Eq. (14) of the potential flow equation $\nabla^2 \phi = 0$, forming the set $[x, y, u^{pf}, v^{pf}, p^{pf}]$, the outputs are the same as PINNs. Both networks contain 5 hidden layers with 64 neurons per layer. PINNs and FENN are trained for 2000 and 500 epochs, respectively. In FENN, $[u^{pf}, v^{pf}, p^{pf}]$ is obtained analytically according to Eq. (14).

$$\phi(x, y) = V_\infty x \left(1 + \frac{R^2}{r^2}\right), r = \sqrt{x^2 + y^2}$$

$$u^{pf} = \frac{\partial \phi}{\partial x}, \quad v^{pf} = \frac{\partial \phi}{\partial y} \qquad (14)$$

$$p^{pf} = \frac{1}{2} V_\infty^2 + p_\infty - \frac{1}{2}[(u^{pf})^2 + (v^{pf})^2]$$

where $\phi$ represents velocity potential, $V_\infty = \sqrt{u_\infty^2 + v_\infty^2}$, $R = d/2 = 0.5$ represents the radius of the cylinder.

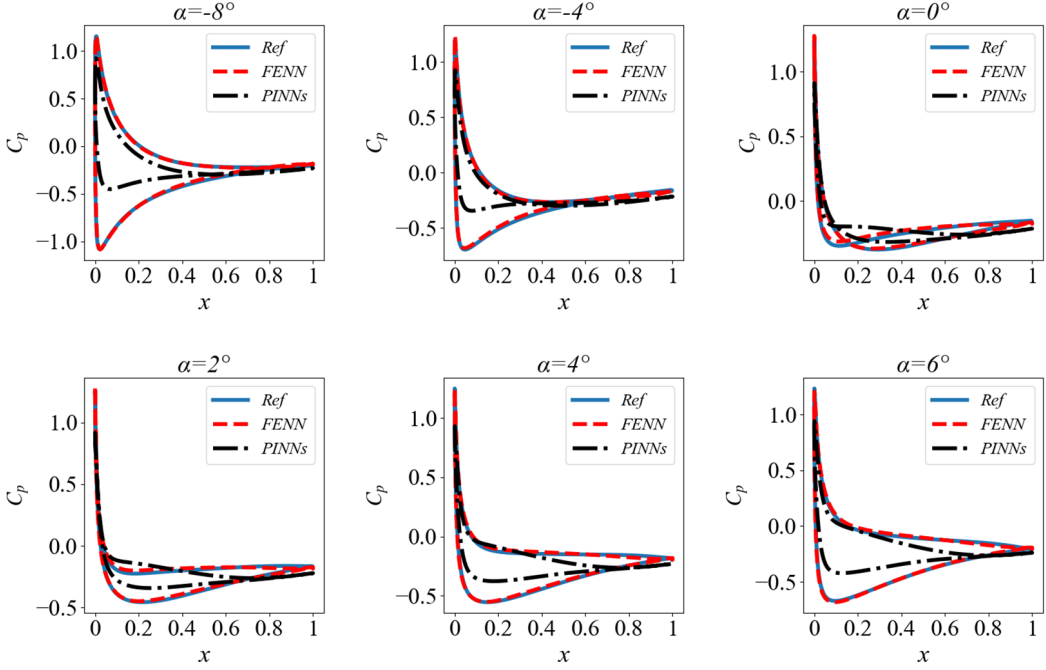

Figure 16. Comparison of the pressure coefficient on the NACA2412 airfoil at different angles of attack obtained by finite volume method (blue), PINNs (black), and FENN (red) for solving parametric problem involving viscous incompressible flow around the NACA2412 airfoil.

Figure 18 shows the convergence history of PINNs and FENN. We observe that compared to PINNs, FENN reduces the relative $L_2$ error of the pressure coefficient on the cylinder to 1% with only one-quarter of the epochs. The corresponding pressure coefficients are shown in Figure 19. The training times for PINNs and FENN are 12.38 minutes and 3.25 minutes, meaning the speedup of FENN is 3.81. In addition, we find that the acceleration effect of physical features is nearly identical to that of geometric features.

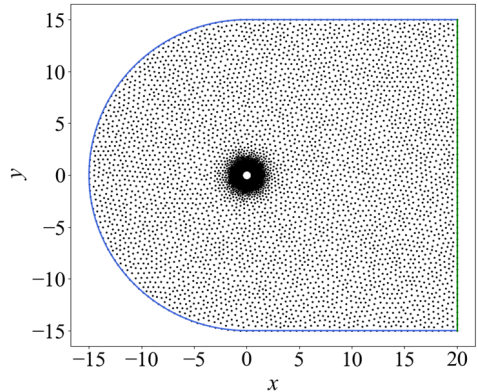

Figure 17. The computational domain and the distribution of collocation points for solving forward problem involving viscous incompressible flow around a circular cylinder.

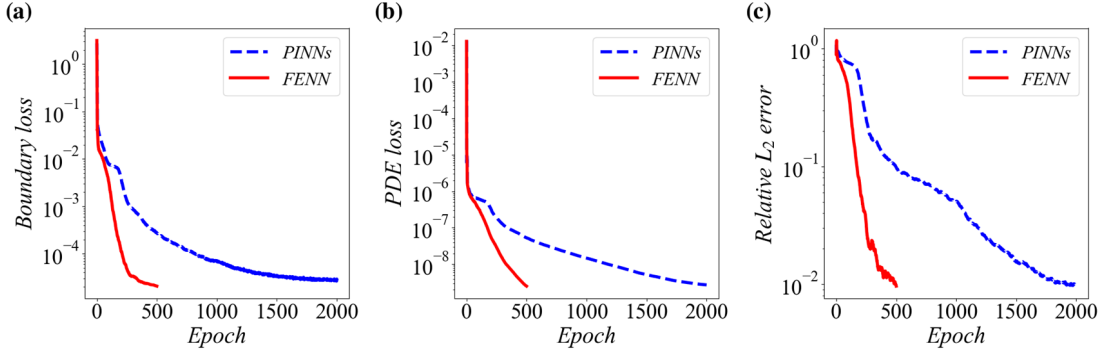

Figure 18. Comparison of the convergence history obtained by PINNs and FENN for solving the forward problem involving viscous incompressible flow around a circular cylinder. (a) Boundary loss. (b) PDE loss. (c) Relative $L_2$ error of the pressure coefficient on the airfoil.

## 4 Conclusions

In this study, we propose the feature-enhanced neural network to solve PDEs involving fluid dynamics. Geometric features including distance and angle or physical features including the solution of the potential flow equation are introduced in the inputs of PINNs, enabling FENN to more easily learn the flow. Consequently, the accuracy and efficiency of solving PDEs involving fluid dynamics are improved. The designed features offer the advantages of low computational cost and high relevance to the problem being solved. To prevent incorrect partial derivatives in FENN due to neglecting the partial derivatives of the features with respect to space-time coordinates, we establish the feature networks to replace the features in the FENN inputs and train them offline in advance.

By solving three forward problems, one inverse problem, and one parametric problem involving different flow conditions and geometries, we comprehensively validate the effectiveness of FENN.

(1) Compared to PINNs, FENN generally reduces the computation time by four times.

(2) FENN obtains high-accuracy results with extremely sparse observed data in solving the inverse problem, whereas PINNs exhibit unacceptable errors.

(3) FENN successfully solves the parametric problem about angle of attack involving the viscous incompressible flow around the NACA2412 airfoil, where PINNs fail.

Some challenges remain, such as the limited generality of the two features we design, as they are specific to fluid dynamics. In the future, more universal features are expected, adaptively designing features through the outputs of the neural network is worth considering.

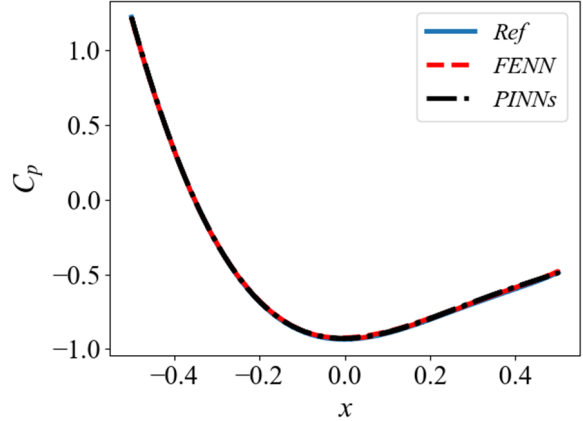

Figure 19. Comparison of the pressure coefficient on the cylinder obtained by finite volume method (blue), PINNs (black), and FENN (red) for solving the forward problem involving viscous incompressible flow around a circular cylinder.

## Acknowledgements

We would like to acknowledge the support of the National Natural Science Foundation of China (No. 92152301).

## References

[1] R. Courant, K. Friedrichs, H. Lewy, Über die partiellen Differenzengleichungen der mathematischen Physik, Mathematische annalen, 100 (1928) 32-74.
[2] R. Eymard, T. Gallouët, R. Herbin, Finite volume methods, Handbook of numerical analysis, 7 (2000) 713-1018.
[3] R. Courant, Variational methods for the solution of problems of equilibrium and vibrations, (1943).
[4] M. Raissi, P. Perdikaris, G.E. Karniadakis, Physics-informed neural networks: A deep learning framework for solving forward and inverse problems involving nonlinear partial differential equations, Journal of Computational Physics, 378 (2019) 686-707.
[5] A.G. Baydin, B.A. Pearlmutter, A.A. Radul, J.M. Siskind, Automatic differentiation in machine learning: a survey, Journal of Marchine Learning Research, 18 (2018) 1-43.
[6] M. Raissi, A. Yazdani, G.E. Karniadakis, Hidden fluid mechanics: Learning velocity and pressure fields from flow visualizations, Science, 367 (2020) 1026-1030.
[7] L. Sun, H. Gao, S. Pan, J.-X. Wang, Surrogate modeling for fluid flows based on physics-constrained deep learning without simulation data, Computer Methods in Applied Mechanics and Engineering, 361 (2020) 112732.
[8] W. Cao, J. Song, W. Zhang, Solving high-dimensional parametric engineering problems for inviscid flow around airfoils based on physics-informed neural networks, Journal of Computational Physics, 516 (2024) 113285.
[9] W. Cao, S. Tang, Q. Ma, W. Ouyang, W. Zhang, Solving all laminar flows around airfoils all-at-once using a parametric neural network solver, arXiv preprint

arXiv:2501.01165, (2025).

[10] X. Jin, S. Cai, H. Li, G.E. Karniadakis, NSFnets (Navier-Stokes flow nets): Physics-informed neural networks for the incompressible Navier-Stokes equations, Journal of Computational Physics, 426 (2021) 109951.

[11] Z. Mao, A.D. Jagtap, G.E. Karniadakis, Physics-informed neural networks for high-speed flows, Computer Methods in Applied Mechanics and Engineering, 360 (2020) 112789.

[12] A.D. Jagtap, Z. Mao, N. Adams, G.E. Karniadakis, Physics-informed neural networks for inverse problems in supersonic flows, Journal of Computational Physics, 466 (2022) 111402.

[13] J. Song, W. Cao, F. Liao, W. Zhang, VW-PINNs: A volume weighting method for PDE residuals in physics-informed neural networks, Acta Mechanica Sinica, 41 (2025) 324140.

[14] W. Cao, J. Song, W. Zhang, A solver for subsonic flow around airfoils based on physics-informed neural networks and mesh transformation, Physics of Fluids, 36 (2024).

[15] J. Wu, Y. Wu, G. Zhang, Y. Zhang, Variable linear transformation improved physics-informed neural networks to solve thin-layer flow problems, Journal of Computational Physics, 500 (2024) 112761.

[16] H. Eivazi, M. Tahani, P. Schlatter, R. Vinuesa, Physics-informed neural networks for solving Reynolds-averaged Navier–Stokes equations, Physics of Fluids, 34 (2022).

[17] F. Pioch, J.H. Harmening, A.M. Müller, F.-J. Peitzmann, D. Schramm, O.e. Moctar, Turbulence Modeling for Physics-Informed Neural Networks: Comparison of Different RANS Models for the Backward-Facing Step Flow, Fluids, 8 (2023) 43.

[18] W. Cao, X. Shan, S. Tang, W. Ouyang, W. Zhang, Solving parametric high-Reynolds-number wall-bounded turbulence around airfoils governed by Reynolds-averaged Navier–Stokes equations using time-stepping-oriented neural network, Physics of Fluids, 37 (2025).

[19] S. Cai, Z. Mao, Z. Wang, M. Yin, G.E. Karniadakis, Physics-informed neural networks (PINNs) for fluid mechanics: A review, Acta Mechanica Sinica, 37 (2021) 1727-1738.

[20] W. Cao, W. Zhang, An analysis and solution of ill-conditioning in physics-informed neural networks, Journal of Computational Physics, 520 (2025) 113494.

[21] M. Raissi, Z. Wang, M.S. Triantafyllou, G.E. Karniadakis, Deep learning of vortex-induced vibrations, Journal of Fluid Mechanics, 861 (2019) 119-137.

[22] C. Cheng, H. Meng, Y.-Z. Li, G.-T. Zhang, Deep learning based on PINN for solving 2 DOF vortex induced vibration of cylinder, Ocean Engineering, 240 (2021) 109932.

[23] J. Blazek, Computational fluid dynamics: principles and applications, Butterworth-Heinemann, 2015.

[24] W. Hackbusch, Multi-grid methods and applications, Springer Science & Business Media, 2013.

[25] E. Turkel, Preconditioning techniques in computational fluid dynamics, Annual Review of Fluid Mechanics, 31 (1999) 385-416.

[26] J.N. Kutz, Deep learning in fluid dynamics, Journal of Fluid Mechanics, 814 (2017) 1-4.
[27] C. Jiang, R. Vinuesa, R. Chen, J. Mi, S. Laima, H. Li, An interpretable framework of data-driven turbulence modeling using deep neural networks, Physics of Fluids, 33 (2021).
[28] X. Guo, W. Li, F. Iorio, Convolutional neural networks for steady flow approximation, in: Proceedings of the 22nd ACM SIGKDD international conference on knowledge discovery and data mining, 2016, pp. 481-490.
[29] S. Bhatnagar, Y. Afshar, S. Pan, K. Duraisamy, S. Kaushik, Prediction of aerodynamic flow fields using convolutional neural networks, Computational Mechanics, 64 (2019) 525-545.
[30] Z. Deng, J. Wang, H. Liu, H. Xie, B. Li, M. Zhang, T. Jia, Y. Zhang, Z. Wang, B. Dong, Prediction of transonic flow over supercritical airfoils using geometric-encoding and deep-learning strategies, Physics of Fluids, 35 (2023).
[31] J. Hu, W. Zhang, Flow field modeling of airfoil based on convolutional neural networks from transform domain perspective, Aerospace Science and Technology, 136 (2023) 108198.
[32] K. Zuo, Z. Ye, L. Zhu, X. Yuan, W. Zhang, CycleMLP++: An efficient and flexible modeling framework for subsonic airfoils, Expert Systems with Applications, 260 (2025) 125455.
[33] L.-W. Chen, N. Thuerey, Towards high-accuracy deep learning inference of compressible turbulent flows over aerofoils, arXiv preprint arXiv:2109.02183, (2021).
[34] K. Zuo, Z. Ye, W. Zhang, X. Yuan, L. Zhu, Fast aerodynamics prediction of laminar airfoils based on deep attention network, Physics of Fluids, 35 (2023).
[35] Y. Yang, Y. Xue, W. Zhao, S. Yao, C. Li, C. Wu, Fast flow field prediction of three-dimensional hypersonic vehicles using an improved Gaussian process regression algorithm, Physics of Fluids, 36 (2024).
[36] W. Zhang, P. Xuhao, K. Jiaqing, W. Xu, Heterogeneous data-driven aerodynamic modeling based on physical feature embedding, Chinese Journal of Aeronautics, 37 (2024) 1-6.
[37] J.-W. Hu, W.-W. Zhang, Mesh-Conv: Convolution operator with mesh resolution independence for flow field modeling, Journal of Computational Physics, 452 (2022) 110896.
[38] J.-Z. Peng, N. Aubry, S. Zhu, Z. Chen, W.-T. Wu, Geometry and boundary condition adaptive data-driven model of fluid flow based on deep convolutional neural networks, Physics of Fluids, 33 (2021).
[39] X. Glorot, Y. Bengio, Understanding the difficulty of training deep feedforward neural networks, in: Proceedings of the thirteenth international conference on artificial intelligence and statistics, JMLR Workshop and Conference Proceedings, 2010, pp. 249-256.
[40] D.C. Liu, J. Nocedal, On the limited memory BFGS method for large scale optimization, Mathematical programming, 45 (1989) 503-528.
[41] H. Gao, L. Sun, J.-X. Wang, PhyGeoNet: Physics-informed geometry-adaptive convolutional neural networks for solving parameterized steady-state PDEs on irregular

domain, Journal of Computational Physics, 428 (2021) 110079.

**Appendix A**

In this section, we provide more details on computing partial derivatives in FENN. Using the feature networks as inputs ensures correct partial derivatives, but this method introduces additional computational cost. Specifically, gradient information needs to be propagated to the feature networks when computing partial derivatives in the main network using automatic differentiation, leading to additional matrix computation. Since this process occurs in every epoch of the main network, the computational cost cannot be ignored.

To address this issue, we compute the partial derivatives of the output $F_j, j=1,...,N$ of the feature networks with respect to their inputs $(\mathbf{x},t)$ before training the main network and reconstruct the features to replace the feature networks in the main network inputs. As an example, consider a two-dimensional steady problem with space coordinates $x, y$. We compute $F_j$, $\partial F_j / \partial x$, $\partial F_j / \partial y$, $\partial^2 F_j / \partial x^2$, $\partial^2 F_j / \partial y^2$ and $\partial^2 F_j / \partial x \partial y$ before training the main network. Next, we separate them from the computational graph of the feature networks to obtain $F_j^{de}$, $\partial F_j^{de} / \partial x$, $\partial F_j^{de} / \partial y$, $\partial^2 F_j^{de} / \partial x^2$, $\partial^2 F_j^{de} / \partial y^2$ and $\partial^2 F_j^{de} / \partial x \partial y$. Finally, we reconstruct the features

$$F_j = F_j^{de} + (x - x^{de}) \frac{\partial F_j^{de}}{\partial x} + \frac{1}{2}(x - x^{de})^2 \frac{\partial^2 F_j^{de}}{\partial x^2} + (y - y^{de}) \frac{\partial F_j^{de}}{\partial y} + \\ \frac{1}{2}(y - y^{de})^2 \frac{\partial^2 F_j^{de}}{\partial y^2} + (x - x^{de})(y - y^{de}) \frac{\partial^2 F_j^{de}}{\partial x \partial y}, \quad j = 1,...,N \tag{A1}$$

In Eq. (A1), $x$ and $y$ represent the space coordinates in the inputs of the main network, $x^{de}$ and $y^{de}$ are the results of separating $x$ and $y$ from the main network computational graph. Obviously, the results of Eq. (A1) are equivalent to the output of the feature networks, and their first and second derivatives with respect to $x, y$ are consistent with the first and second derivatives obtained by the feature networks. Therefore, the correct partial derivatives in the main network are still guaranteed, while avoiding the propagation of gradient information in the feature networks.

**Appendix B**

Figure B1 shows the loss convergence of the feature networks in all numerical experiments from Section 3.

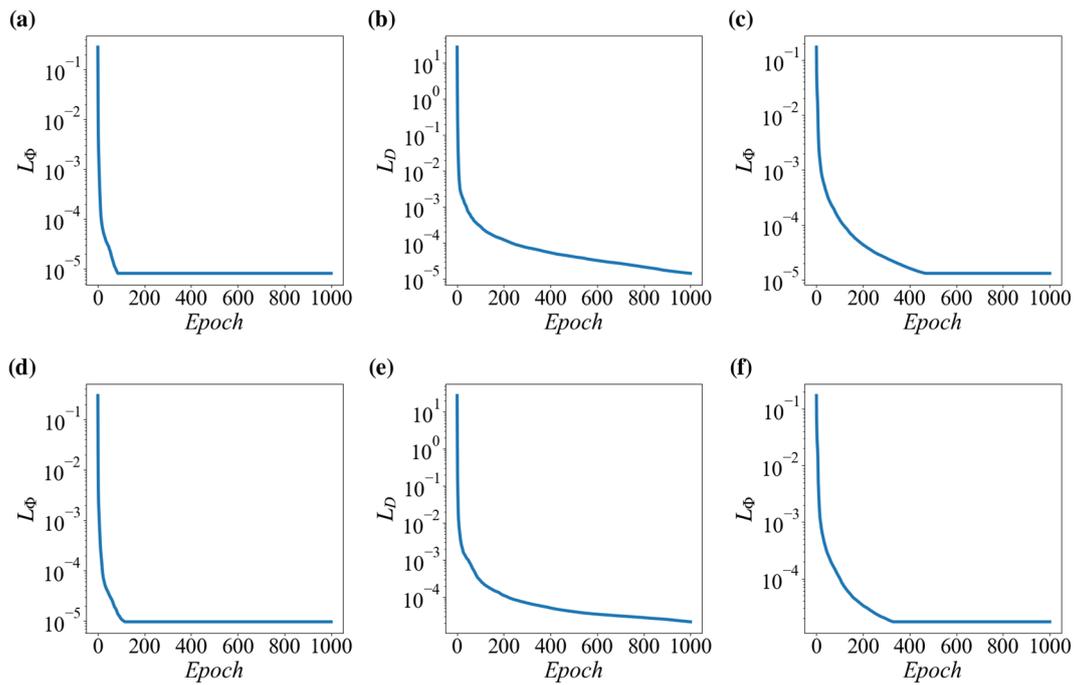

Figure B1. Loss function of the feature networks. (a) corresponding to Section 3.1. (b) and (c) to Section 3.2. (d) to Section 3.3. (e) and (f) to Section 3.4.